\documentclass[preprintnumbers,amsmath,11pt,amssymb,floatfix,superscriptaddress,nofootinbib]{article}

\def\mytitle#1{\setcounter{equation}{0}
\setcounter{footnote}{0}
\begin{flushleft}\Large\textbf{#1}\end{flushleft}
\vspace{0.25cm}}
\def\myname#1{\leftline{{\large #1}}\vspace{-0.13cm}}
\def\myplace#1#2{\small\begin{flushleft}\textit{#1}\\
\texttt{#2}\end{flushleft}}

\usepackage{graphicx}
\begin{document}

\mytitle{Interacting Generalised Cosmic Chaplygin gas in Loop
quantum cosmology: A singularity free universe}

\vskip0.2cm \myname{ Ratul
Chowdhury\footnote{r.chowdhury.024@gmail.com}~$*$} \myplace{$*$
Department of Chemical Engineering, Jadavpur University,
Kolkata-700 032, India.} {} \vskip0.2cm \myname{ Prabir
Rudra~\footnote{prudra.math@gmail.com}$\dag$}
\myplace{$\dag$Department of Mathematics, Bengal Engineering and
Science University, Shibpur, Howrah-711 103, India.} {}

\begin{abstract}
In this work we investigate the background dynamics when dark
energy is coupled to dark matter with a suitable interaction in
the universe described by Loop quantum cosmology. Dark energy in
the form of Generalised Cosmic Chaplygin gas is considered. A
suitable interaction between dark energy and dark matter is taken
into account in order to at least alleviate (if not solve) the
cosmic coincidence problem. The dynamical system of equations is
solved numerically and a stable scaling solution is obtained. A
significant attempt towards the solution of the cosmic coincidence
problem is taken. The statefinder parameters are also calculated
to classify the dark energy model. Graphs and phase diagrams are
drawn to study the variations of these parameters. It is seen that
the background dynamics of Generalised Cosmic Chaplygin gas is
completely consistent with the notion of an accelerated expansion
in the late universe. From the graphs, generalised cosmic
Chaplygin gas is identified as a dark fluid with a lesser negative
pressure compared to Modified Chaplygin gas, thus supporting a 'No
Big Rip' cosmology. It has also been shown that in this model the
universe follows the power law form of expansion around the
critical point, which is consistent with the known results. Future
singularities that may be formed in this model as an ultimate fate
of the universe has been studied in detail. It was found that the
model is completely free from any types of future singularities.
\end{abstract}

\section{Introduction}
At the turn of the last century observations from Ia Supernova
confirmed that our universe is suffering from an accelerated
expansion \cite{Perlmutter1, Spergel1}. In the quest of finding a
suitable model for the expanding universe, Cosmologists started to
investigate the root cause that is triggering this expansion.
Fundamentally, we were to modify Einstein's equation either by
modifying the left hand side, i.e., modifying the idea of Einstein
gravity or to modify the right hand side which immediately
speculate the nature of the matter inside the universe. If our
Universe is filled up by some invisible fluid having a large
negative pressure then it violates the strong energy condition
i.e. $\rho+3p<0$. Due to its invisible nature this energy
component is aptly termed as dark energy (DE) \cite{Riess1}.

With the introduction of DE,  the search began for different
candidates that can effectively play the role of DE.  DE
represented by a scalar field \footnote{ in the presence of a
scalar field the transition from a universe filled with matter to
an exponentially expanding universe is justified } is often called
quintessence. Not only scalar field but also there are other Dark
fluid models like Chaplygin gas which plays the role of DE very
efficiently.  The earliest form of this was known as pure
Chaplygin gas \cite{Kamenshchik1, Gorini1}. Extensive research saw
pure Chaplygin gas modify into generalized Chaplygin gas
\cite{Gorini2, Alam1, Bento1, Carturan1, Barreiro1}. Finally
modified Chaplygin gas (MCG) came into existence \cite{Benaoum1,
Debnath1}. Dynamics of MCG in Loop quantum cosmology was studied
by Jamil et al \cite{Jamil1}. Dynamics of MCG in Braneworld was
studied by Rudra et al \cite{Rudra1}.

Currently, we live in a special epoch where the densities of DE
and dark matter(DM) are comparable, although they have evolved
independently from different mass scales. This is known as the
famous cosmic coincidence problem. till date several attempts have
been made to find a solution to this problem \cite{del Campo1,
Leon1, Jimenez1, Berger1, Zhang1, Griest1, Jamil2, Jamil3, Jamil4,
Jamil5, Jamil6}. A suitable interaction between DE and DM is the
best possible tool to find an effective solution to this problem.
It is obvious that there has been a transition from a matter
dominated universe to dark energy dominated universe, by exchange
of energy at an appropriate rate. Now the expansion history of the
universe as determined by the supernovae and CMB data
\cite{Jamil2, Jamil3} bounds us to fix the decay rate in such a
way that it is proportional to the present day Hubble parameter.
Keeping the above fact in mind cosmologists all over the world
have studied and proposed a variety of interacting DE models
\cite{Setare1, Setare2,  Hu1, Wu1, Jamil7, Jamil8, Dalal1}.

In 2003, P. F. Gonz아lez-Diaz \cite{Gonz아lez-Diaz1} introduced
the generalized cosmic Chaplygin gas (GCCG) model. The speciality
of the model being that it can be made stable and free from
unphysical behaviours even when the vacuum fluid satisfies the
phantom energy condition. In the previous studies related to DE
corresponding to phantom era Big-Rip is essential, as the time
gradient of scale-factor blows to infinity in finite time. For the
first time P. F. Gonz아lez-Diaz including the GCCG model showed
that Big Rip, i.e., singularity at a finite time is totally out of
question. Hence in such models there is no requirement for
evaporation of black hole to zero mass. The Equation of state
(EOS) of this model is
\begin{equation}\label{ratul7.1}
p=-\rho^{-\alpha}\left[C+\left\{\rho^{(1+\alpha)}-C\right\}^{-\omega}\right]
\end{equation}
where $C=\frac{A}{1+\omega}-1$, with A being a constant that can
take on both positive and negative values, and $-{\cal
L}>\omega>0$, ${\cal L}$ being a positive definite constant, which
can take on values larger than unity. GCCG can explain the
evolution of the universe starting from the dust era to $\Lambda
CDM$, radiation era, matter dominated quintessence and lastly
phantom era \cite{Chakraborty1}.

As we have stated earlier, modifying the right hand side of
Einstein's equation was not the only way to explain the increase
in the rate of expansion of the universe. The gravity part on the
left hand side of Einstein's equation can also be modified to
demonstrate the present day universe. In this perspective Loop
quantum gravity (LQG) was introduced and Loop quantum cosmology
(LQC) was developed. In this work we consider LQG as the modified
gravity theory. Our main aim of this work is to examine the nature
of the different physical parameters for the universe around the
stable critical points in LQC model in presence of GCCG. Impact of
any future singularity caused by the DE in LQC model will be
studied as well.

This paper is organized as follows: Section 2 comprises of the
basic equations of the LQC model. Section3 deals with the
dynamical system analysis of GCCG in LQC model. In Section 4 we
present a complete graphical representation of phase plane
analysis. In Section 5 we study the future singularities, followed
by some concluding remarks in section 6.

\section{The Model: Loop Quantum Gravity}
In recent years, loop Quantum Gravity (LQG) has evolved as an
outstanding effort to describe the quantum effect of our universe
\cite{Rovelli1, Ashtekar1}. LQG is a theory trying to quantize the
gravity with a non-perturbative and background independent method.
The theory and principles of LQG when applied in the cosmological
framework creates a new theoretical framework of Loop Quantum
Cosmology(LQC) \cite{Ashtekar2, Bojowald1, Ashtekar3}. In this
theory, classical space-time continuum is replaced by a discrete
quantum geometry. The effect of LQG can be described by the
modification of Friedmann equation by adding a term quadratic in
density. In LQC, the non-perturbative effects lead to correction
term $\frac{\rho_{T}^{2}}{\rho_{1}}$ to the standard Friedmann
equation. With the inclusion of this term, the universe bounces
quantum mechanically as the matter energy density reaches the
level of $\rho_{1}$(order of Plank density).

\subsection{Basic equations of Loop quantum gravity model}\label{sec2}
Recently the model of DE has been explored in the framework of
LQC. The cosmological evolution in LQC has been widely studied for
quintessence and phantom DE models \cite{Wu2}. The modified
Friedmann equation for Loop Quantum Cosmology is given by
\cite{Wu1,Chen1,Fu1}
\begin{equation}\label{ratul7.2}
H^2=\frac{\rho_{T}}{3}\left(1-\frac{\rho_{T}}{\rho_{1}}\right)
\end{equation}
Here $\rho_{1}=\sqrt{3}\pi^{2}\gamma^{3}G^{2}\hbar$ is the
critical loop quantum density and $\gamma$ is the dimensionless
Barbero-Immirzi parameter. $\rho_{T}=\rho_{m}+\rho_{E}$ represents
the total cosmic energy density, which is a sum of energy density
of DM (${\rho}_{m}$) and the energy density of DE ($\rho_{E}$).

As in the present problem the interaction between DE in the form
of GCCG and pressureless DM has been taken into account. For
interacting GCCG and DM the energy balance equation will be
\begin{equation}\label{ratul7.3}
\dot{\rho}_{gccg}+3H\left(1+w_{gccg}\right)\rho_{gccg}=-Q,~~~for
~GCCG~
\end{equation}
and ~
\begin{equation}\label{ratul7.4}
\dot{\rho}_m+3H\rho_m=Q, ~for~ the~ DM~ interacting ~with~ GCCG.
\end{equation}
where $Q=3bH\rho$ is the interaction term, $b$ is the coupling
parameter (or transfer strength) and
$\rho_{T}=\rho=\rho_{gccg}+\rho_m$ is the total cosmic energy
density which satisfies the energy conservation equation
$\dot{\rho}+3H\left(\rho+p\right)=0$ \cite{Guo1, del Campo1}.

As we have lack of information about the fact, how DE and DM
interact so we are not able to estimate the interaction term from
the first principles. However, the negativity of $Q$ immediately
implies the possibility of having negative DE in the early
universe which is overruled by the necessity of the second law of
thermodynamics to be held \cite{Alcaniz1}. Hence $Q$ must be
positive and small. From the observational data of 182 Gold type
Ia supernova samples, CMB data from the three year WMAP survey and
the baryonic acoustic oscillations from the Sloan Digital Sky
Survey, it is estimated that the coupling parameter between DM and
DE must be a small positive value (of the order of unity), which
satisfies the requirement for solving the cosmic coincidence
problem and the second law of thermodynamics \cite{Feng1}. Due to
the underlying interaction, the beginning of the accelerated
expansion is shifted to higher redshifts. Consequently using the
Friedmann equation (\ref{ratul7.2}) and the conservation equation,
we obtain the modified Raychaudhuri equation
\begin{equation}\label{ratul7.5}
\dot{H}=-\frac{1}{2}\left(\rho+p\right)\left(1-2\frac{\rho}{\rho_{1}}\right)
\end{equation}

\section{Dynamical system analysis}
In this section we plan to analyse the dynamical system. For that
firstly we convert the physical parameters into some dimensionless
form, given by
\begin{equation}\label{ratul7.6}
x=\ln a, ~~~~~~~ u=\frac{\rho_{gccg}}{3H^2},
~~~~~~~v=\frac{\rho_m}{3H^2}
\end{equation}
where the present value of the scale factor, $a_0=1$ is assumed.
Using equation (\ref{ratul7.1}) to equation (\ref{ratul7.6}) we
get the parameter gradients as given below:
\begin{equation}\label{ratul7.7}
\frac{du}{dx}=3\left[u\left(\frac{1-v}{1+v}\right)\left\{v+u\left(1+w_{gccg}\right)\right\}-b\left(u+v\right)-u\left(1+w_{gccg}\right)\right]
\end{equation}
and
\begin{equation}\label{ratul7.8}
\frac{dv}{dx}=3\left[b\left(u+v\right)-v+v\left(\frac{1-v}{1+v}\right)\left\{v+u\left(1+w_{gccg}\right)\right\}\right]
\end{equation}
Where, $w_{gccg}$ is the EoS parameter for GCCG determined as
\begin{equation}\label{ratul7.9}
w_{gccg}=\frac{p_{gccg}}{\rho_{gccg}}=\frac{\left(u+v\right)^{4}\left[-C-\left(-C+\frac{u^{2}\left(u+v-1\right)^{2}\rho_{1}^{2}}{\left(u+v\right)^{4}}\right)^{-\omega}\right]}{u^{2}\left(u+v-1\right)^{2}\rho_{1}^{2}}
\end{equation}
In the above calculations for mathematical simplicity we have
considered $\alpha=1$.

\subsection{Critical points}
The critical points of the above system are obtained by putting
$\frac{du}{dx}=0=\frac{dv}{dx}$.  Now this system of equation does
not yield an explicit solution. So we have to investigate for a
numerical solution by putting some special values to the
parameters which will yield a non zero positive solution of the
system of equations. Considering $\omega=-1$, we obtain the
following critical values of the system.

\begin{equation}\label{ratul7.10}
u_{1c}=\frac{1}{2b}\left(\frac{1}{4}+\frac{1}{4}P-\frac{Q}{2\sqrt{2}}\right)\left[\frac{7}{4}-\frac{1}{4}P-3b+\frac{Q}{2\sqrt{2}}-b\left(\frac{1}{4}+\frac{1}{4}P-\frac{Q}{2\sqrt{2}}\right)+2\left(\frac{1}{4}+\frac{1}{4}P-\frac{Q}{2\sqrt{2}}\right)^{2}-\left(\frac{1}{4}+\frac{1}{4}P-\frac{Q}{2\sqrt{2}}\right)^{3}\right]
\end{equation}
\begin{equation}\label{ratul7.11}
v_{1c}=\frac{1}{4}+\frac{1}{4}P-\frac{Q}{2\sqrt{2}}
\end{equation}
and
\begin{equation}\label{ratul7.12}
u_{2c}=\frac{1}{2b}\left(\frac{1}{4}+\frac{1}{4}P+\frac{Q}{2\sqrt{2}}\right)\left[\frac{7}{4}-\frac{1}{4}P-3b-\frac{Q}{2\sqrt{2}}-b\left(\frac{1}{4}+\frac{1}{4}P+\frac{Q}{2\sqrt{2}}\right)+2\left(\frac{1}{4}+\frac{1}{4}P+\frac{Q}{2\sqrt{2}}\right)^{2}-\left(\frac{1}{4}+\frac{1}{4}P+\frac{Q}{2\sqrt{2}}\right)^{3}\right]
\end{equation}
\begin{equation}\label{ratul7.13}
v_{2c}=\frac{1}{4}+\frac{1}{4}P+\frac{Q}{2\sqrt{2}}
\end{equation}
where~~~~ $P=\sqrt{1-4b}$ ~~~~~~~and~~~~~~
$Q=\sqrt{-3+\frac{5}{\sqrt{1-4b}}-2b-\frac{20b}{\sqrt{1-4b}}}$

The above two critical points correspond to the era dominated by
DM and GCCG type DE and exist for $b\leq\frac{1}{4}$. For the
critical points the equation of state parameter (\ref{ratul7.9})
of the interacting DE takes the form
\begin{equation}\label{ratul7.14}
w_{gccg}=\frac{\left(u_{ic}+v_{ic}\right)^{4}\left[-C-\left(-C+\frac{u_{ic}^{2}\left(u_{ic}+v_{ic}-1\right)^{2}\rho_{1}^{2}}{\left(u_{ic}+v_{ic}\right)^{4}}\right)^{-\omega}\right]}{u_{ic}^{2}\left(u_{ic}+v_{ic}-1\right)^{2}\rho_{1}^{2}}
\end{equation}
where $i=1,~2$ ~~~and which holds for~~~ $u_{ic}+v_{ic}\neq1$.

\subsection{Stability around critical point}
Now we check the stability of the dynamical system (eqs.
(\ref{ratul7.7}) and (\ref{ratul7.8})) about the critical points.
In order to do this, we linearize the governing equations about
the critical points i.e.,
\begin{equation}\label{ratul7.15}
u=u_c+\delta u ~~  and ~~  v=v_c+\delta v,
\end{equation}
Now if we assume $f=\frac{du}{dx}$ and $g=\frac{dv}{dx}$, then we
may obtain
\begin{equation}\label{ratul7.16}
\delta\left(\frac{du}{dx}\right)=\left[\partial_{u}
f\right]_{c}\delta u+\left[\partial_{v} f\right]_{c}\delta v
\end{equation}
and
\begin{equation}\label{ratul7.17}
\delta\left(\frac{dv}{dx}\right)=\left[\partial_{u}
g\right]_{c}\delta u+\left[\partial_{v} g\right]_{c}\delta v
\end{equation}

where
$$\partial_{u}{f}=\left[3\left\{-CB^{2}M^{3}\left(\left(4+b-4u\right)\left(u-1\right)u+\left(u+u^{2}\left(4u-11\right)+b\left(u^{2}+u-1\right)\right)v+\left(1+u\left(7u+2b-5\right)\right)v^{2}\right.\right.\right.$$
$$\left.\left.\left.+\left(4u+b-2\right)v^{3}+v^{4}\right)+\frac{1}{u^{2}\rho_{1}^{2}}N^{-\omega}\left(-u^{2}B^{2}M^{3}\left(2u^{3}\left(v-1\right)-3u\left(v+1\right)-2u\left(1+v\left(3+\left(v-1\right)v\right)\right)\omega\right.\right.\right.\right.$$
$$\left.\left.\left.\left.-v\left(v^{2}-1\right)\left(1+2\omega\right)+u^{2}\left(5+v\left(-5-2v\left(\omega-1\right)\right)\rho_{1}^{2}+u^{4}B^{5}\left(1+2u\left(v-1\right)+v^{2}+b\left(1+v\right)\right)\rho_{1}^{4}N^{\omega}\right.\right.\right.\right.\right.$$
\begin{equation}\label{ratul7.18}
\left.\left.\left.\left.\left.+CM^{7}\left(2u^{3}\left(v-1\right)+v-v^{3}-3u\left(v+1\right)+u^{2}\left(5+v\left(2v-5\right)\right)\right)\left(1+CN^{\omega}\right)\right)\right)\right)\frac{1}{\left(v+1\right)B^{3}\left(CM^{4}-u^{2}B^{2}\rho_{1}^{2}\right)}\right\}\right]
\end{equation}

$$\partial_{v}{f}=\frac{1}{\left(1+v\right)^{2}}3N^{-\omega}\left[-\frac{2u^{2}M^{4}}{CM^{4}-u^{2}B^{2}\rho_{1}^{2}}+\frac{1}{-CuB^{3}M^{4}\rho_{1}^{2}+u^{3}B^{5}\rho_{1}^{4}}\left\{-u^{3}B^{5}\left(b\left(v+1\right)^{2}+u\left(-1+2u+v\left(2+v\right)\right)\right)\right.\right.$$
$$\left.\left.\left.\rho_{1}^{4}N^{w}-2CM^{7}\left(-2+u^{3}-3v+v^{3}+u\left(3+v^{3}\right)+u^{2}\left(-2+v\left(2+v\right)\right)\right)\left(1+CN^{\omega}\right)+uB^{2}M^{3}\rho_{1}^{2}\left(2u\left(v+1\right)\left(u+v-2\right)\right.\right.\right.\right.$$
$$\left.\left.\left.\left.\left(1+u\left(v-1\right)+v\right)\left(1+\omega\right)+C\left(4u^{4}+b\left(v-1\right)v\left(v+1\right)^{2}+u^{3}\left(v\left(3v+10\right)-7\right)+u^{2}\left(7+b\left(v+1\right)^{2}+v\left(v+2\right)\left(4v-3\right)\right)\right.\right.\right.\right.\right.$$
\begin{equation}\label{ratul7.19}
\left.\left.\left.\left.+u\left(-4+b\left(v+1\right)^{2}\left(2v-1\right)+v\left(v+1\right)\left(-5+v\left(v+2\right)\right)\right)\right)N^{\omega}\right)\right\}\right]
\end{equation}

$$\partial_{u}{g}=3\left[b+\left\{v\left(v-1\right)N^{-\omega}\left(-u^{4}B^{5}\rho_{1}^{4}N^{\omega}-CM^{7}\left(u\left(u-3\right)+v-v^{2}\right)\left(1+CN^{\omega}\right)-u^{2}B^{2}M^{3}\rho_{1}^{2}\left(-3u+u^{2}+v-v^{2}\right.\right.\right.\right.$$
\begin{equation}\label{ratul7.20}
\left.\left.\left.\left.-2u\omega+2v\omega-2uv\omega-2v^{2}\omega+2Cu\left(u+v-2\right)N^{\omega}\right)\right)\right\}\frac{1}{u^{2}\left(1+v\right)B^{3}\rho_{1}^{2}\left(-CM^{4}+u^{2}B^{2}\rho_{1}^{2}\right)}\right]
\end{equation}

$$\partial_{v}{g}=3\left[1+b-u-2v+\frac{2\left(u-1\right)}{\left(1+v\right)^{2}}+\left\{M^{3}N^{-\omega}\left(CM^{4}\left(v\left(v-1\right)\left(3v^{2}-5\right)+u^{2}\left(v\left(v+2\right)-1\right)+u\left(1+v\left(v\left(4v+3\right)\right.\right.\right.\right.\right.\right.$$
$$\left.\left.\left.\left.\left.\left.-6\right)\right)\right)\left(1+CN^{\omega}\right)-u^{2}B^{2}\rho_{1}^{2}\left(u-u^{2}+5v-6uv+2u^{2}v-5v^{2}+3uv^{2}+u^{2}v^{2}-3v^{3}+4uv^{3}+3v^{4}+4v\omega-2uv\omega-2v^{2}\omega\right.\right.\right.\right.$$
$$\left.\left.\left.\left.-4v^{3}\omega+2v^{4}\omega+C\left(v\left(v-1\right)\left(3v^{2}-5\right)+u^{2}\left(v\left(v+2\right)-1+u\left(1+v\left(-6+v\left(4v+3\right)\right)\right)\right)N^{\omega}\right)\right)\right)\right.\right.$$
\begin{equation}\label{ratul7.21}
\left.\left.\frac{1}{\left(1+v\right)^{2}B^{3}\rho_{1}^{2}\left(CuM^{4}-u^{3}B^{2}\rho_{1}^{2}\right)}\right\}\right]
\end{equation}
where $B=u+v-1$, ~~~$M=u+v$,
~~~$N=-C+\frac{u^{2}\left(u+v-1\right)^{2}\rho_{1}^{2}}{\left(u+v\right)^{4}}$

\subsection{Nature of cosmological parameters}
We calculate the deceleration parameter ~$q=-1-(\dot H/H^2)$, in
this model as,
\begin{equation}\label{ratul7.22}
q=-1+\frac{3}{2}\left(1+\omega_{gccg}\frac{\rho_{gccg}}{\rho}\right)\left(\frac{1-2\frac{\rho}{\rho_{1}}}{1-\frac{\rho}{\rho_{1}}}\right)
\end{equation}
Now expressing the above expression in the form of dimensionless
density parameter $\Omega_{gccg}=\frac{\rho_{gccg}}{\rho}$, we
get,

$$q=-1+\frac{3}{2}\left(1+\omega_{gccg}\Omega_{gccg}\right)\left(\frac{1-2\frac{\rho}{\rho_{1}}}{1-\frac{\rho}{\rho_{1}}}\right)$$
From the above expression it is evident that in the limit
$\rho_{1}\rightarrow\infty$, we retrieve the result for Einstein's
gravity as follows,

$$q_{EG}=-1+\frac{3}{2}\left(1+\omega_{gccg}\Omega_{gccg}\right)$$
 Now assuming $\frac{\rho}{\rho_{1}}=\epsilon \sim O(1)$ and using
 the transformations of equation (\ref{ratul7.6}), we get,
\begin{equation}\label{ratul7.23}
q=-1+\frac{3}{2}\left(1+\frac{u\omega_{gccg}}{u+v}\right)\left(\frac{1-2\epsilon}{1-\epsilon}\right)
\end{equation}
Since a physically acceptable solution corresponds to both the
critical points, such that $\left(u,v\right)\rightarrow
\left(u_{ic},v_{ic}\right), ~~~i=1,2$. Hence from equation
(\ref{ratul7.23}) we get,
\begin{equation}\label{ratul7.24}
q_{c}=-1+\frac{3}{2}Z, ~~~~where
~~~~~~Z=\left(1+\frac{u_{ic}\omega_{gccg}}{u_{ic}+v_{ic}}\right)\left(\frac{1-2\epsilon}{1-\epsilon}\right),~~~~~i=1,
2
\end{equation}

As special cases, we observe that for $\epsilon = \frac{1}{2}$, we
have $q =-1$ while $\epsilon=1$, we have $q=-\infty$
 if $\omega_{gccg}\geq-\frac{1}{u_{1c}}$.

 We know that the Hubble parameter varies as
\begin{equation}\label{ratul7.25}
H=\frac{2}{3Zt}
\end{equation}
where we have ignored the integration constant. Integration of the
above equation (\ref{ratul7.25}) yields,
\begin{equation}\label{ratul7.26}
a\left(t\right)=a_{0}t^{\frac{2}{3Z}}
\end{equation}
which, as its form suggests gives a power law form of expansion.

Sahni et al \cite{Sahni1} introduced a pair of cosmological
diagnostic pair $\{r,s\}$ which is known as statefinder
parameters. The two parameters are dimensionless and are
geometrical since they are derived from the cosmic scale factor
alone. Also this pair generalizes the well-known geometrical
parameters like the Hubble parameter and the deceleration
parameter. The statefinder parameters are given by
\begin{equation}\label{27}
r\equiv\frac{\stackrel{...}a}{aH^3},\ \
s\equiv\frac{r-1}{3(q-1/2)}.
\end{equation}
In the LQC model, we have the following expressions of $r$ and $s$
as
\begin{equation}\label{28}
r_{(LQC)}=\left(1-\frac{3Z}{2}\right)\left(1-3Z\right).
\end{equation}
and
\begin{equation}\label{29}
s_{(LQC)}=Z.
\end{equation}
It is interesting to note that the pair $\{r_{LQC}, s_{LQC}\}$
yields the $\Lambda CDM$ (cosmological constant-cold dark matter
model) $\{r_{EG}, s_{EG}\} = \{1, 0\}$ when $Z = 0$ or
$\epsilon=\frac{1}{2}$.

\begin{figure}
\includegraphics[height=1.5in]{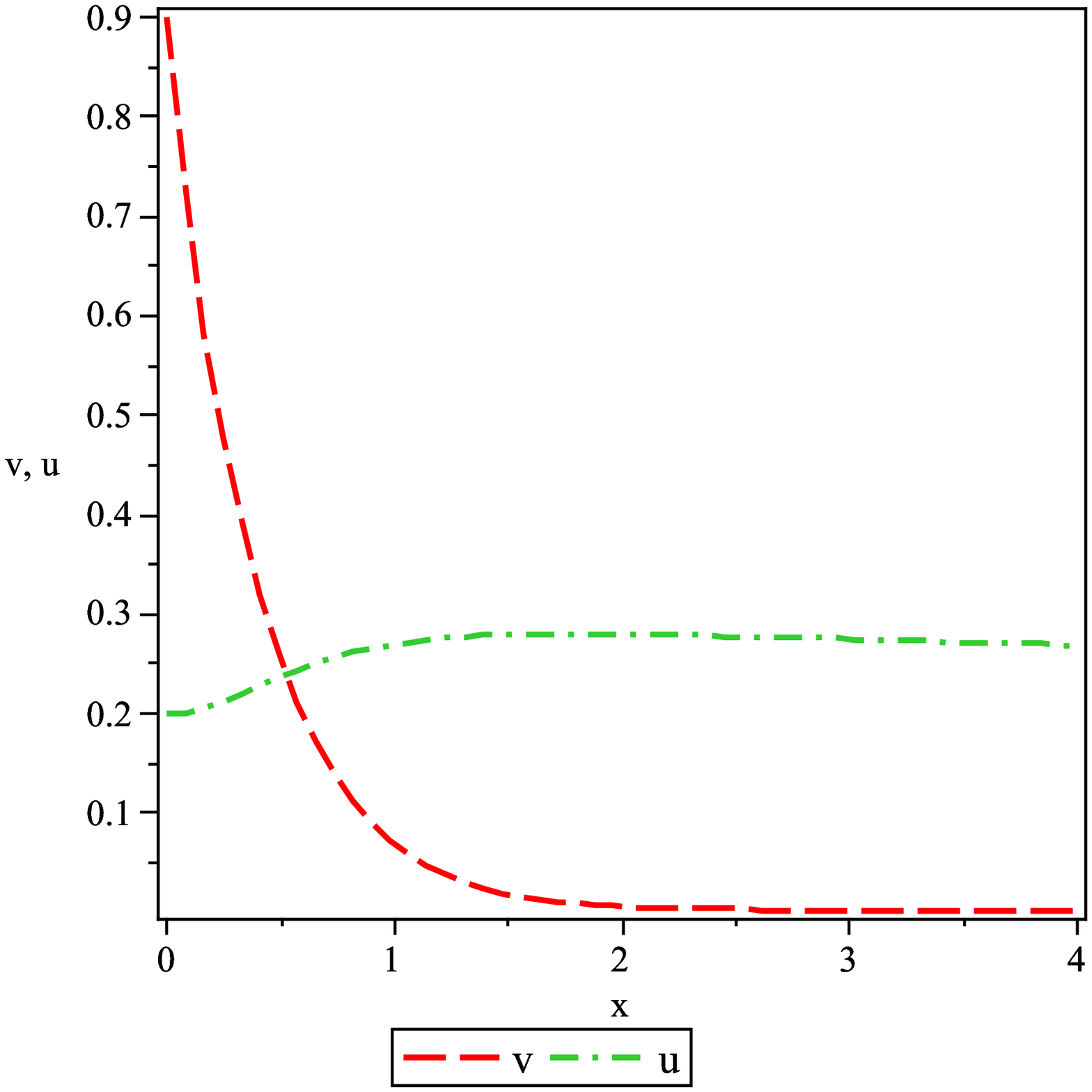}~~~~~~~~\includegraphics[height=1.5in]{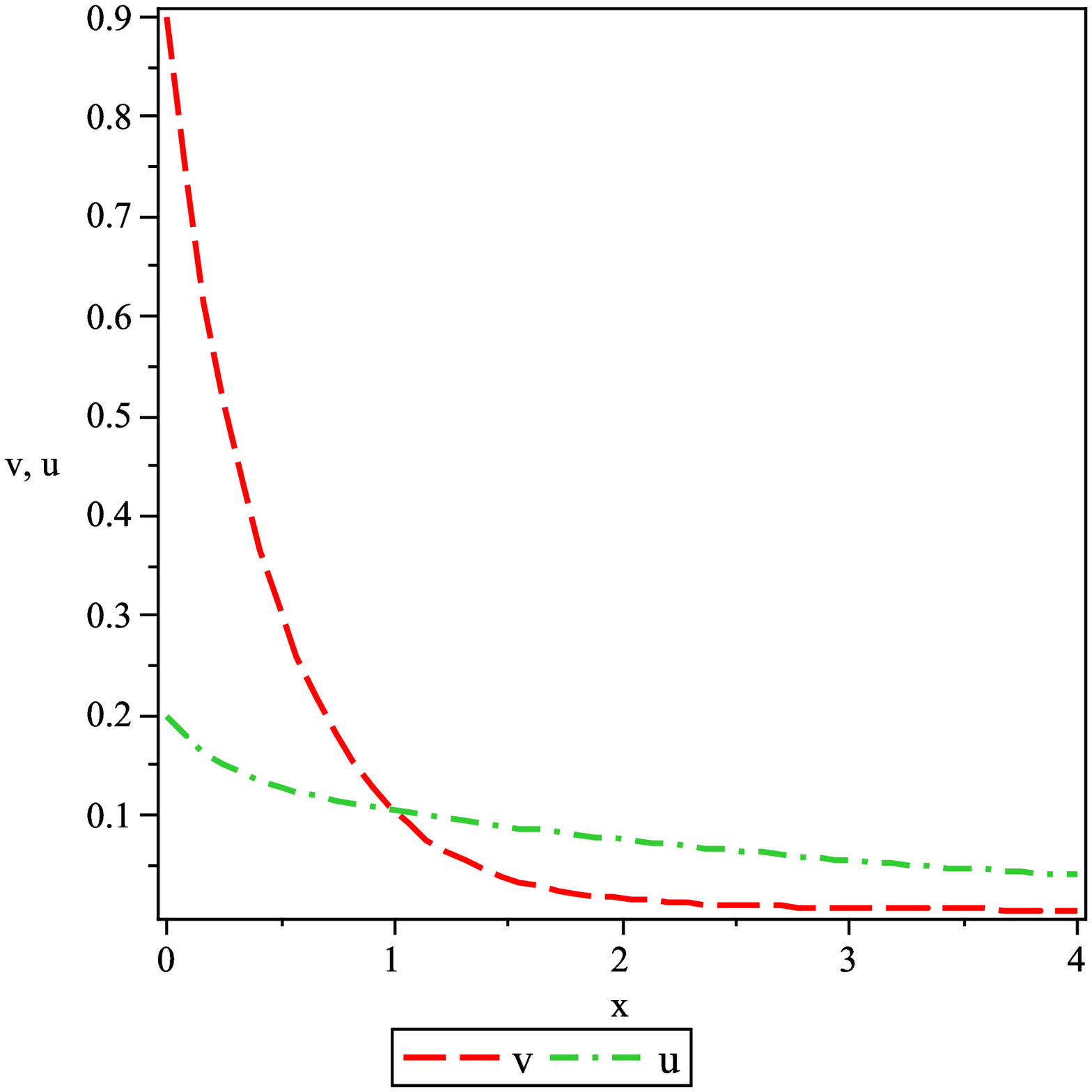}~~~~~~~~\includegraphics[height=1.5in]{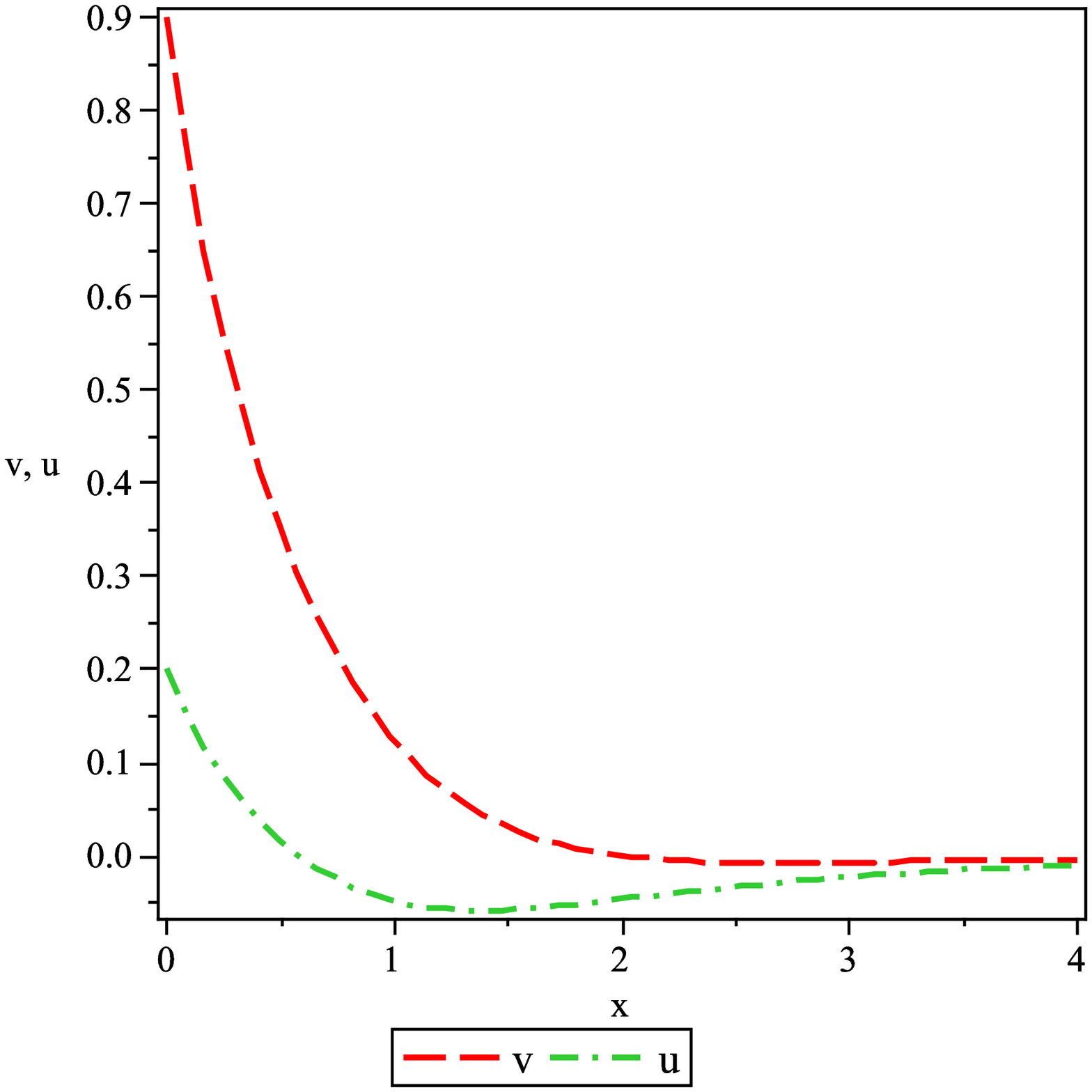}~~~~~~~\includegraphics[height=1.5in]{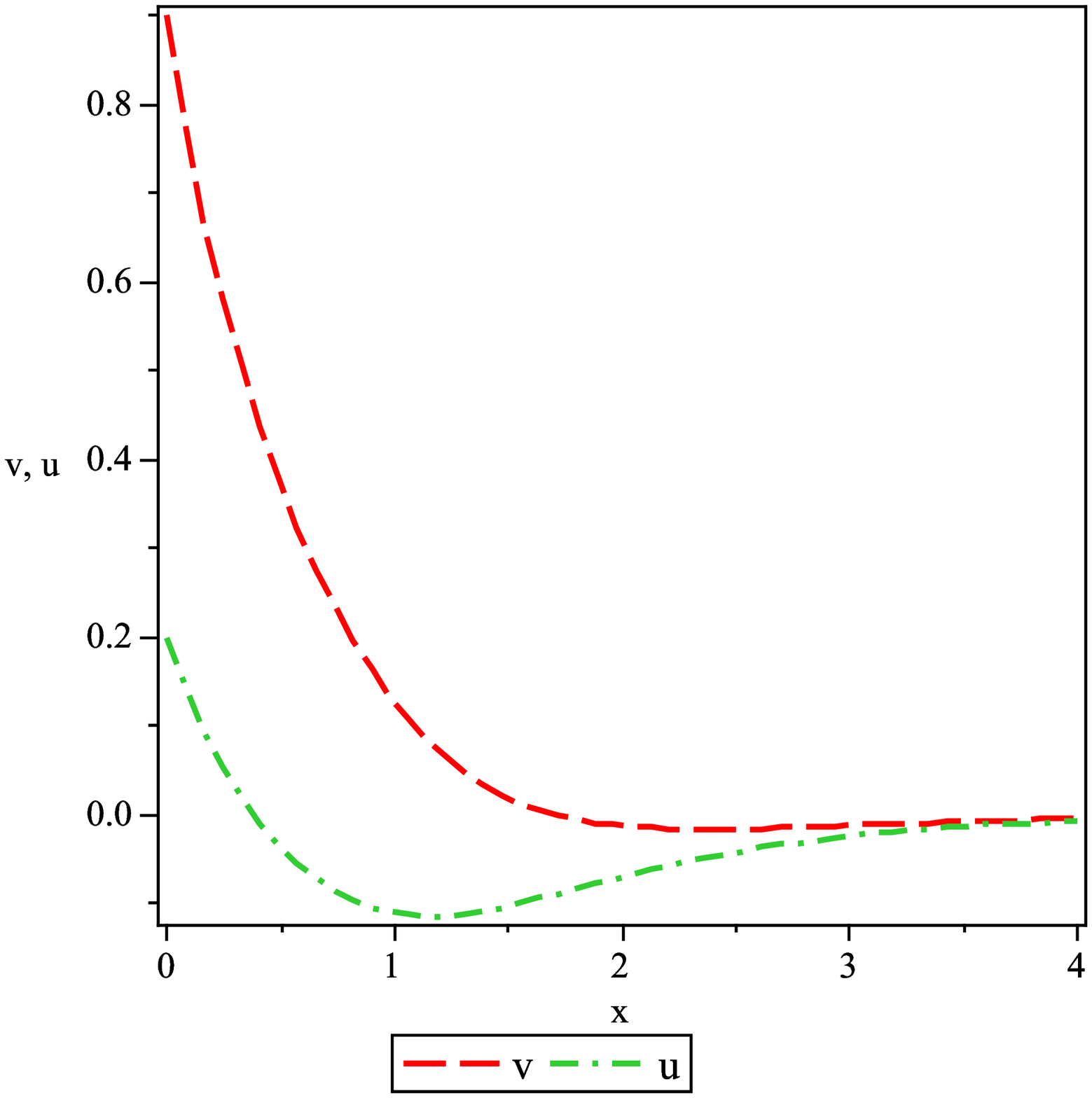}~\\
\vspace{1mm}
~~~~~~~~~~~~Fig. 1~~~~~~~~~~~~~~~~~~~~~~~~Fig. 1a~~~~~~~~~~~~~~~~~~~~~~~~~~~~~Fig. 1b~~~~~~~~~~~~~~~~~~~~~~~~~~Fig. 1c\\
\vspace{1mm} Figs.1, 1a, 1b, 1c : The dimensionless density
parameters are plotted against e-folding time. The initial
condition is $v(0)=0.9, u(0)=0.2$. The other parameters are fixed
at $\omega=-1, C=1$ and $\rho_{1}=10$.
The interactions are respectively $b=0.01, 0.1, 0.2$ and $0.25$\\

\includegraphics[height=2in]{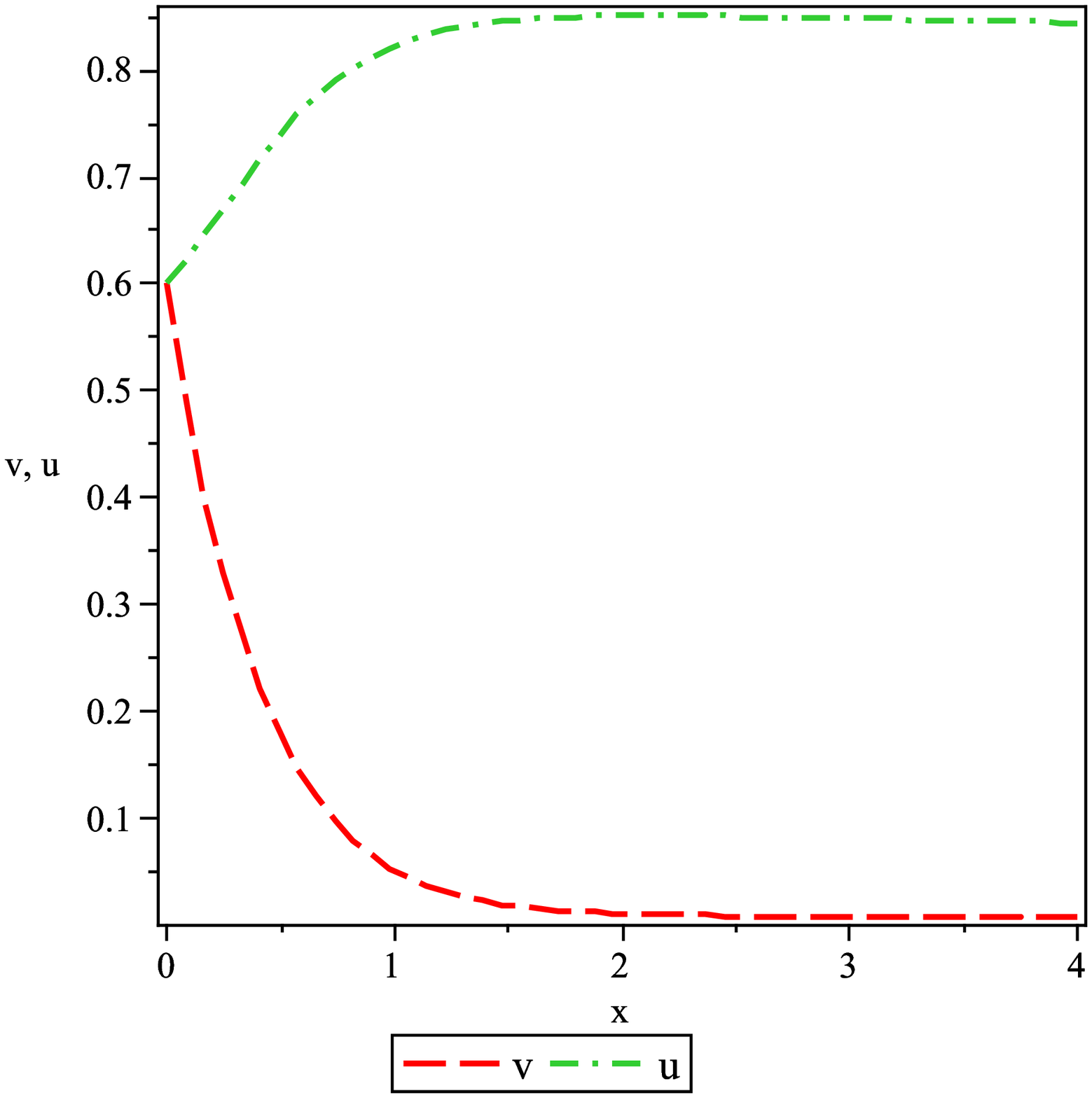}~~~~~\includegraphics[height=2in]{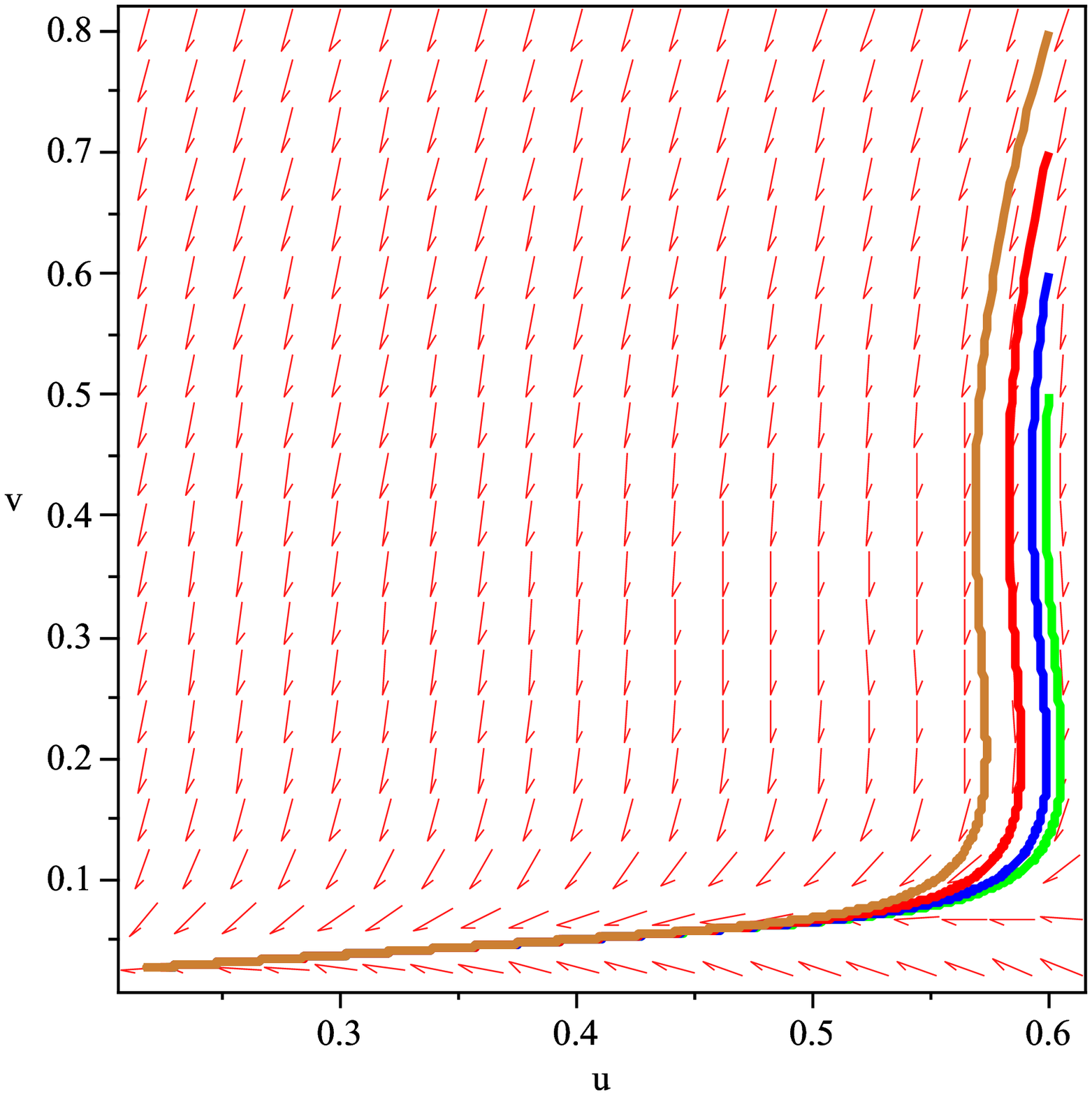}
~~~~~\includegraphics[height=2in]{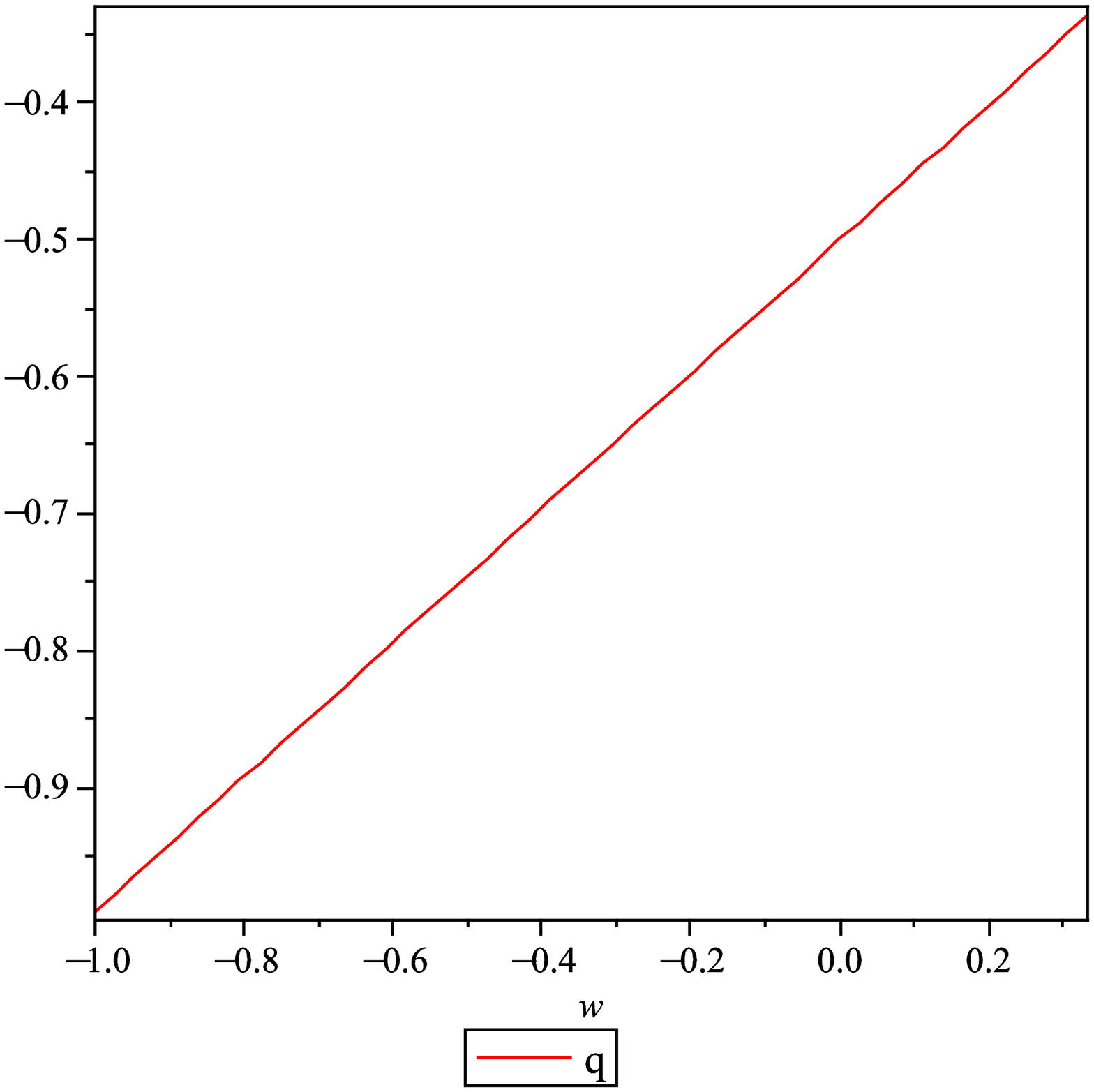}~\\
\vspace{1mm}
~~~~~~~~~~~~~Fig. 2~~~~~~~~~~~~~~~~~~~~~~~~~~~~~~~~~~~~~~~~Fig. 3~~~~~~~~~~~~~~~~~~~~~~~~~~~~~~~~~~Fig. 4\\
\vspace{1mm} Fig 2 : The dimensionless density parameters are
plotted against e-folding time.The initial condition is $v(0)=0.6,
u(0)=0.6$. The other parameters are fixed at $\omega=-1, b=0.01,
C=1$ and $\rho_{1}=10$.\\
Fig 3 :  The phase diagram of the parameters depicting an
attractor solution. The initial conditions chosen are $v(0)=0.5,
u(0)=0.6$ (green); $v(0)=0.6, u(0)=0.6$ (blue); $v(0)=0.7,
u(0)=0.6$ (red); $v(0)=0.8, u(0)=0.6$ (brown). Other parameters
are fixed at $\omega=-1, b=0.1, C=1$ and $\rho_{1}=10$.\\
 Fig 4 :The deceleration parameter is plotted against the state parameter.
Other parameters are fixed at $\omega=-1, b=0.01, C=1,
\rho_{1}=10, \epsilon= 0.4$.

\includegraphics[height=2in]{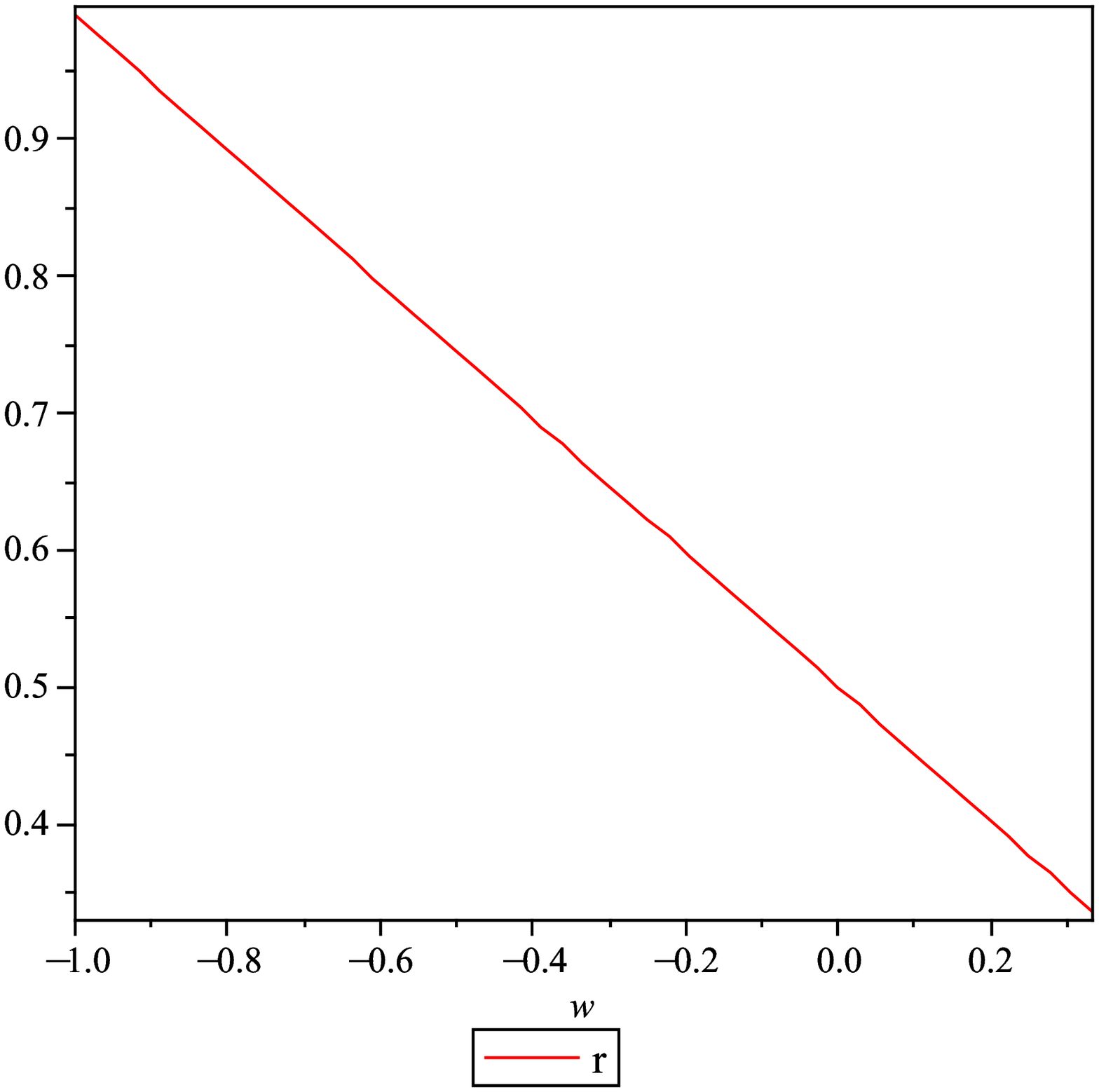}~~~~~\includegraphics[height=2in]{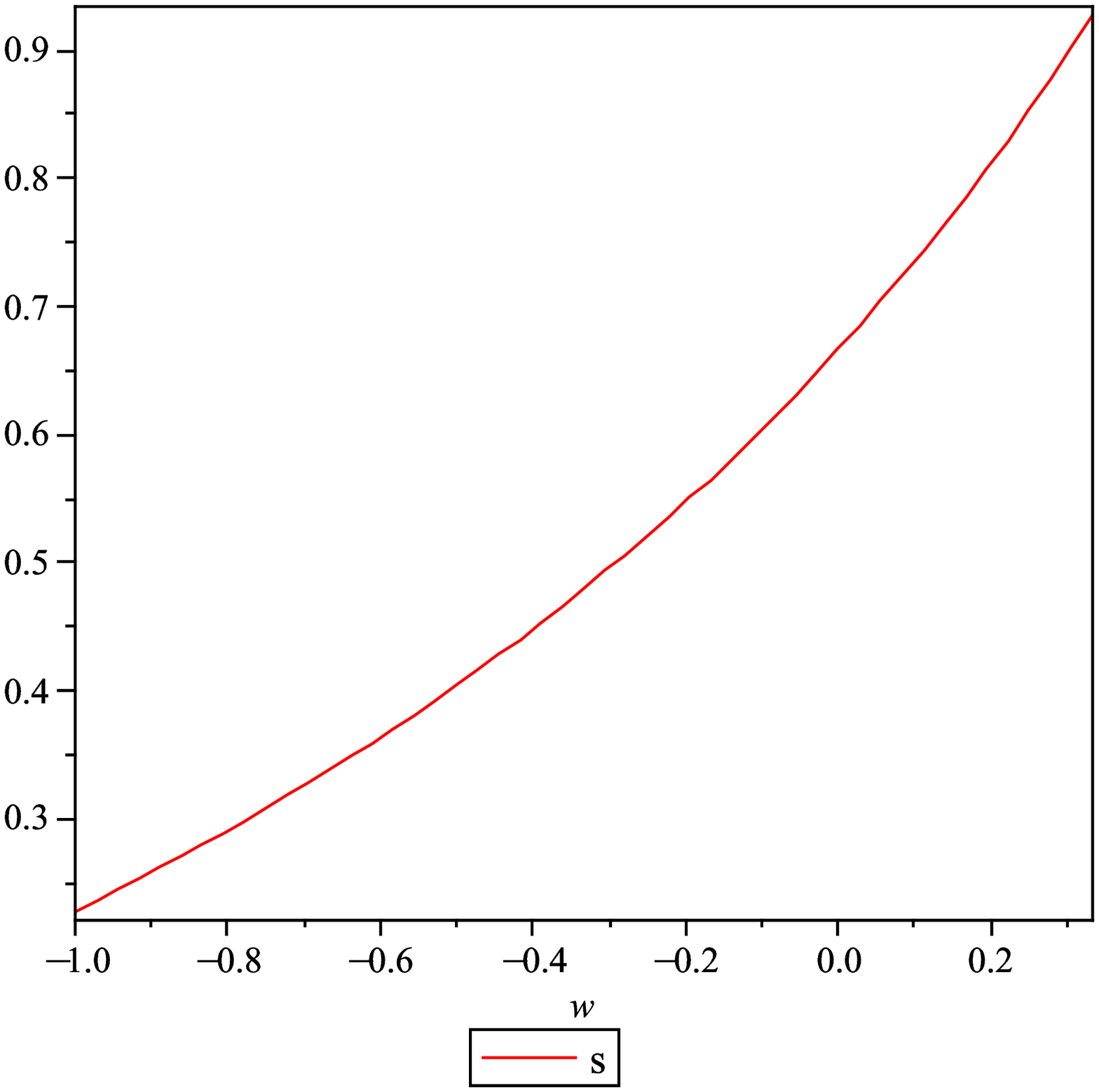}~~~~~~~~~~~~\includegraphics[height=2in]{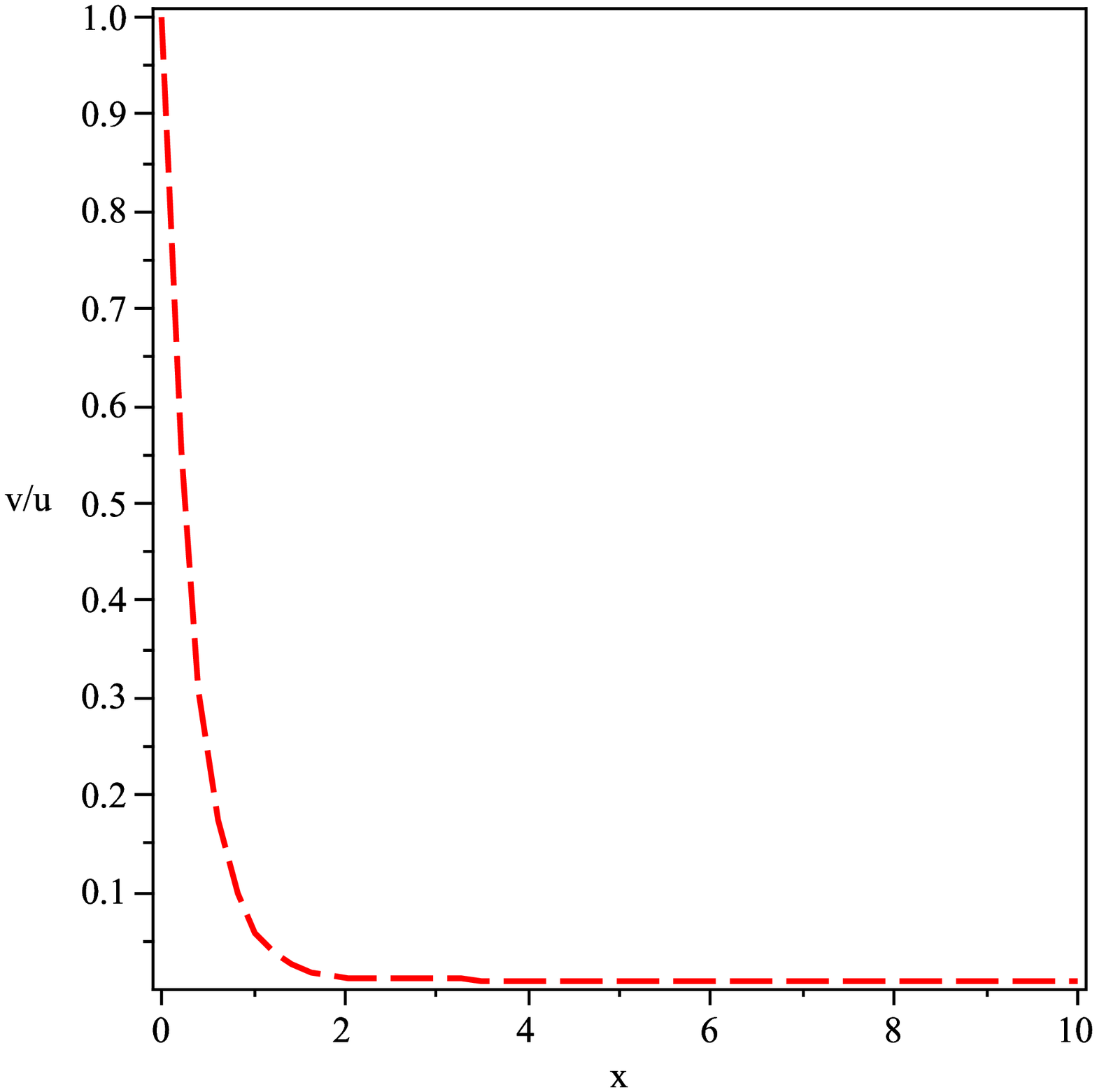}~\\
\vspace{1mm}
~~~~~~~~~~Fig. 5~~~~~~~~~~~~~~~~~~~~~~~~~~~~~~~~~~~~~~~~Fig. 6~~~~~~~~~~~~~~~~~~~~~~~~~~~~~~~~~~~~~~~~~~~~~~Fig. 7\\
\vspace{1mm} Fig 5 :  The statefinder parameter $r$ is plotted
against the state parameter $\omega_{gccg}$. The other parameters
are fixed at $\omega=-1, b=0.01, C=1, \rho_{1}=10, \epsilon= 0.4$.\\
Fig 6 : The statefinder parameter $s$ is plotted against the state
parameter $\omega_{gccg}$. The other parameters are fixed at
$\omega=-1, b=0.01, C=1, \rho_{1}=10, \epsilon= 0.4$.\\
Fig 7 : The ratio of density parameters is shown against e-folding
time. The initial conditions chosen are v(0)=0.6, u(0)=0.6. The
other parameters are fixed at $\omega=-1, b=0.01, C=1, \rho_{1}=10$.\\
\end{figure}

\section{Graphical Representation of the Phase plane Analysis}
Phase diagrams are drawn to determine the type of critical point
obtained in this model. We discuss the results obtained in detail
below:

The dimensionless density parameters $v$ and $u$ are drawn in
figures 1, 1a, 1b, 1c and 2. From the figures we see that $v$
decreases, and $u$ increases during evolution of the universe.
This shows that the density of DM decreases while the density of
DE increases as the universe evolves. So this result is consistent
with the well known idea of an energy dominated universe. Figs 1,
1a, 1b and 1c are graphs depicting the same quantities with
gradually increasing magnitude of interaction between GCCG and DM.
It is clearly evident from the figures that with the increase in
interaction between GCCG and DM, their respective density
parameters ($u$ and $v$) become more and more comparable to each
other. Thus this interacting model probably gives the best
solution to the cosmic coincidence problem. Moreover from the
figs.1, 1a, 1b, 1c and 2, it is seen that the graph of DE (green
curve) does not show any urgency to creep up steadily in the
vertical direction. This is a striking difference from the graphs
obtained by Jamil et al \cite{Jamil1} and Rudra et al
\cite{Rudra1} in case of MCG in LQC and MCG in Braneworld
respectively. So in this case it is understandable that for GCCG,
the the relative domination of the DE component over DM is far
lesser as compared to MCG. The possible reason may be a weaker
negative pressure of GCCG compared to that of MCG. This may well
be the underlying reason for GCCG supporting a 'No Big Rip'
cosmology. The phase space diagram (figure 3) shows the attractor
solution. The eigen values are calculated at both the critical
points. At the first critical point eigen values are found to be
$(0.0381055,~ 1.52435)$. So the critical point is an unstable
node. At the second critical point the eigen values are
$(0.0381047,~ -6.09814)$. Hence the critical
point is a saddle point.\\

Figure 4 shows the variations of the deceleration parameter, $q$
against $\omega_{gccg}$. From figure 4, it is evident that there
is a gradual decrease in the deceleration parameter $q$, and
finally in the late universe it attains negative values, which
suggests that there should be an acceleration in the late
universe. This result is in accordance with the various
observational data of Ia supernovae and CMB data which suggests
that the universe is undergoing an accelerated expansion of late.
Figure 5 and 6 shows the variations of the statefinder parameters
$r$ and $s$ against $\omega_{gccg}$. From these figures it can be
seen that $r$ tends towards $1$ and $s$ tends towards $0$.
Therefore it is evident that these results tends towards the
$\Lambda CDM$ model. Figure 7 shows the variation of the ratio of
the density parameters $v$ and $u$ against time. From the figure
it is evident that the ratio of the above parameters decreases
with time. So it can be concluded that there is a relative
decrease in matter density with respect to the energy density.
This is again consistent with the notion of an energy dominated
universe.

\section{Study of Future Singularities}
It is a well known fact that any energy dominated model of the
universe is destined to result in a future singularity. The study
of dynamics of an accelerating universe in the presence of DE and
DM is in fact incomplete without the study of these singularities,
which are the ultimate fate of the universe.  It is known that the
universe dominated by phantom energy ends with a future
singularity known as Big Rip \cite{Caldwell1}, due to the
violation of dominant energy condition (DEC). But other than this
there are other types of singularities as well. Nojiri et al
\cite{Nojiri1} studied the various types of singularities that can
result from an phantom energy dominated universe. These possible
singularities are characterized by the growth of energy and
curvature at the time of occurrence of the singularity. It is
found that near the singularity quantum effects becomes very
dominant which may alleviate or even prevent these singularities.
So it is extremely necessary to study these singularities and
classify them accordingly so that we can search for methods to
eliminate them. The appearance of all four types of future
singularities in coupled fluid dark energy, $F(R)$ theory,
modified Gauss-Bonnet gravity and modified $F(R)$ Horava-Lifshitz
gravity was demonstrated in \cite{Nojiri2}. The universal
procedure for resolving such singularities that may lead to bad
phenomenological consequences was proposed. In Rudra et al
\cite{Rudra1} it has been shown that in case of Modified Chaplygin
gas(MCG), both Type I and Type II singularities are possible.

\subsection{TYPE I Singularity (Big Rip singularity)}
If $\rho\rightarrow\infty$ , $|p|\rightarrow\infty$ when
$a\rightarrow\infty$ and $t\rightarrow t_{s}$. Then the
singularity formed is said to be the Type I singularity.

In the present case by considering the GCCG equation of state from
equation (\ref{ratul7.1}) we find that there is no possibility for
TypeI singularity, i.e., Big Rip singularity. This is in absolute
accordance with P. F. Gonz아lez-Diaz who has successfully shown
that by considering GCCG as the DE, Big Rip can easily be avoided,
thus giving a singularity free late universe.

\subsection{TYPE II Singularity (Sudden singularity)}
If $\rho\rightarrow\rho_{s}$ and $\rho_{s}\sim0$, then
$|p|\rightarrow-\infty$ for $t\rightarrow t_{s}$ and $a\rightarrow
a_{s}$, then the resulting singularity is called the Type II
singularity.

In this case we consider the equation of state for GCCG, like the
previous case for our investigation. We see that if
$\rho\rightarrow\rho_{s}$ and $\rho_{s}\sim0$, then
$|p|\rightarrow0$ for $t\rightarrow t_{s}$ and $a\rightarrow
a_{s}$. Hence there is no possibility of the type II singularity
or the sudden singularity in case of GCCG.

\subsection{TYPE III Singularity}
For $t\rightarrow t_{s}$, $a\rightarrow a_{s}$,
$\rho\rightarrow\infty$ and $|p|\rightarrow\infty$. Then the
resulting singularity is Type III singularity. It is quite evident
from the equation of state of GCCG that it does not support this
type of singularity.

\subsection{TYPE IV Singularity}
For $t\rightarrow t_{s}$, $a\rightarrow a_{s}$, $\rho\rightarrow0$
and $|p|\rightarrow0$. Then the resulting singularity is Type IV
singularity. This type of singularity is not supported by GCCG
type DE.

As a remark, one should stress that our consideration is totally
classical. Nevertheless, it is expected that quantum gravity
effects may play significant role near the singularity. It is
clear that such effects may contribute to the singularity
occurrence or removal too. Unfortunately, due to the absence of a
complete quantum gravity theory only preliminary estimations may
be done.

\section{Concluding Remarks}

In this work, we have considered a combination of Generalised
cosmic Chaplygin gas in Loop quantum gravity model. Our basic idea
was to study the background dynamics of GCCG in detail when it is
incorporated in Loop quantum gravity. Dynamical system analysis
had been carried out, critical points were found and the stability
of the system around those critical points was tested. Graphical
analysis was done to get an explicit picture of the outcome of the
work. In order to find a solution for the cosmic coincidence
problem, a suitable interaction between DE and DM was considered.
Figures of density parameters were drawn for different values of
interaction. It was found that increase in interaction resulted in
more and more comparable values of the density parameters of GCCG
and DM. Since the tendency of DE domination over DM is lesser in
case of GCCG compared to MCG, GCCG is identified as a dark fluid
with a lesser negative pressure compared to MCG. The dynamical
system of equations characterizing the system was formed and a
stable scaling solution was obtained. Hence this work can be
considered to be a significant one, if not the best, as far as the
solution of cosmic coincidence problem is concerned. Study of
future singularities had been carried out in detail. The model was
investigated for all possible types of future singularities. From
the above analysis we conclude that the combination of GCCG in
loop quantum gravity gives a perfect
singularity free model for an expanding universe undergoing a late acceleration. \\

\end{document}